\documentclass[trackchanges, twocolumn]{aastex7}
\usepackage{float}
\usepackage[utf8]{inputenc}

\usepackage{graphicx}
\usepackage{subcaption}
\usepackage{booktabs} 
\usepackage{amsmath}

\begin{document}

\title{Can GRB Empirical Correlations Be Used for Population Studies? }

\author[orcid=0000-0002-8442-9458,sname='North America']{Emre S. Yorgancioglu}
\affiliation{Key Laboratory of Particle Astrophysics, Institute of High Energy Physics, Chinese Academy of Sciences \\
19B Yuquan Road, Beijing 100049, People’s Republic of China}
\affiliation{University of Chinese Academy of Sciences, Chinese Academy of Sciences, Beijing 100049, People’s Republic of China \\
}

\email[show]{emre@ihep.ac.cn}  

\author[orcid=0000-0002-1905-1727]{Yun-Fei Du}

\affiliation{Key Laboratory of Particle Astrophysics, Institute of High Energy Physics, Chinese Academy of Sciences \\
19B Yuquan Road, Beijing 100049, People’s Republic of China}
\affiliation{University of Chinese Academy of Sciences, Chinese Academy of Sciences, Beijing 100049, People’s Republic of China \\
}
\email{fakeemail3@google.com}

\author[orcid=0000-0001-7599-0174,gname=Bosque, sname='Sur America']{ Shu-Xu Yi} 
\affiliation{Key Laboratory of Particle Astrophysics, Institute of High Energy Physics, Chinese Academy of Sciences \\
19B Yuquan Road, Beijing 100049, People’s Republic of China}
\email[show]{sxyi@ihep.ac.cn}

\author[orcid=0000-0001-5586-1017]{Shuang-Nan Zhang}
\affiliation{Key Laboratory of Particle Astrophysics, Institute of High Energy Physics, Chinese Academy of Sciences \\
19B Yuquan Road, Beijing 100049, People’s Republic of China}
\email[show]{zhangsn@ihep.ac.cn}

%\collaboration{all}{The Terra Mater collaboration}

%% Use the \collaboration command to identify collaborations. This command
%% takes an optional argument that is either a number or the word "all"
%% which tells the compiler how many of the authors above the command to
%% show. For example "\collaboration[all]{(DELVE Collaboration)}" wil include
%% all the authors above this command.
%%
%% Mark off the abstract in the ``abstract'' environment. 
\begin{abstract}

\noindent Only a small fraction of Gamma-ray bursts (GRBs) have independent redshift measurements, which are essential for understanding their intrinsic properties. For this reason, empirical correlations of GRBs have often been touted as useful distance indicators, for both individual GRBs as well as population studies. Building upon our previous work, we test the ability of the Yonetoku, 3D Dainotti, and the $L$–$T$–$E$ correlation to adequately constrain the GRB rate, \( \Psi_{\mathrm{GRB}} \).  Our analysis demonstrates that, even under idealized conditions that neglect substantial uncertainties, the derived redshift solutions cannot accurately constrain \( \Psi_{\mathrm{GRB}} \), regardless of the intrinsic distribution's characteristic width. \textcolor{black}{We thus demonstrate unequivocally that—notwithstanding the questionable assumption of no selection biases—empirical GRB correlations alone cannot serve as reliable distance indicators at either the individual or population level.}

%$E_{\mathrm{p},z} - L_{\mathrm{p},z}$,  $L_{a} -T_{a}-L_{p}$ and  $L_{a} -T_{a}-E_{\mathrm{iso}}$,

%Previously, we demonstrated the challenges in estimating pseudo-redshifts of Gamma-ray bursts (GRBs) using two phenomenological correlations of the prompt emission: the Amati and Yonetoku relations. While the Yonetoku relation provides a unique redshift solution, its intrinsic data dispersion results in larger systematic errors that often surpass the typical width of the redshift distribution.  We find that even when these substantial systematic errors are disregarded, the derived redshift solutions cannot accurately constrain the simulated GRB formation rate, \( \Psi_{\mathrm{GRB}} \), regardless of the intrinsic distribution's characteristic width. We also test the case for a 3 parameter correlations of $L_{a} -T_{a}-L_{p}$ and  $L_{a} -T_{a}-E_{\mathrm{iso}}$, and show conclusively that empirical correlations   
\end{abstract}

%% Keywords should appear after the \end{abstract} command. 
%% The AAS Journals now uses Unified Astronomy Thesaurus (UAT) concepts:
%% https://astrothesaurus.org
%% You will be asked to selected these concepts during the submission process
%% but this old "keyword" functionality is maintained in case authors want
%% to include these concepts in their preprints.
%%
%% You can use the \uat command to link your UAT concepts back its source.

%% From the front matter, we move on to the body of the paper.
%% Sections are demarcated by \section and \subsection, respectively.
%% Observe the use of the LaTeX \label
%% command after the \subsection to give a symbolic KEY to the
%% subsection for cross-referencing in a \ref command.
%% You can use LaTeX's \ref and \label commands to keep track of
%% cross-references to sections, equations, tables, and figures.
%% That way, if you change the order of any elements, LaTeX will
%% automatically renumber them.

\section{Introduction} 
\noindent Gamma-ray bursts (GRBs) are the brightest transient phenomenon in the universe, with the brightest of all time GRB 221009A reaching an unprecedented isotropic-equivalent energy of $10^{56}$ ergs \citep{Frederiks2023}. Their extreme energetics, combined with their extra-galactic origin have prompted interest in the possibility of their being used for cosmological studies \citep{2008Amati}. GRBs have traditionally been categorized based on the duration of their prompt emission, with Short Gamma-ray Bursts (SGRBs) defined with $T_{90} < 2$s, and Long Gamma-ray Bursts (LGRBs) defined with $T_{90} > 2$s. It has been of the belief that SGRBs result from the merger of compact objects, while LGRBs are the result of collapsars. However, observations have increasingly revealed cases where GRBs exhibit physical characteristics that do not align neatly with the expectations derived solely from their durations \citep{norris2006short, 2025ApJ...979...73W}. Drawing an analogy with the classification of supernovae, \citet{zhang2006burst} proposed a physically motivated scheme that distinguishes GRBs based on their progenitor systems—categorizing them as either Type I (typically associated with compact object mergers) or Type II (linked to the core-collapse of massive stars), taking into account multi-wavelength observations and host galaxy properties.

\noindent GRBs have been shown to exhibit varied empirical correlations in both the prompt and afterglow phases; \cite{Amati2002} found a relation between the intrinsic spectral peak energy $E_{\mathrm{p}, z}$ (in the $\nu f_{\nu} $ spectrum) and the total equivalent isotropic energy $E_{\mathrm{iso}}$, known as the ``Amati" relation. Similarly, \cite{yonetoku2004gamma} found a relation between $E_{\mathrm{p}, z}$ and the isotropic peak luminosity $L_{\mathrm{p}, z}$, ie, 

\begin{equation}
\log\left(\frac{E_{\text{p},z}}{\text{keV}}\right) = a_{\rm Y} \log\left(\frac{L_{\text{p},z}}{\text{erg/s}}\right) + b_{\rm Y}  \,,
\end{equation}

\noindent Among the most well-known correlations of the afterglow was discovered by \cite{Dainotti2008},  which links the isotropic plateau luminosity $L_X(T^{*}_a) = L_X$ and the rest-frame time at which the plateau ends in the X-ray afterglows of GRBs $T_a^*$, with the relation $\log L_X =  a + b \log T_a^*$, known as the Dainotti relation. This was later extended to a three parameter correlation with the inclusion of $L_{\mathrm{p}, z}$, linking the prompt and afterglow, in the form of a fundamental plane: 

\begin{equation}
\log L_X = C_0 + \alpha \log L_{\text{p}, z} + \beta \log T^{*}_a
\label{Eq3DD}
\end{equation}

\noindent with  \( C_0 = 15.75 \pm 5.3 \), \( \alpha = 0.67 \pm 0.1 \), and \( \beta = -0.77 \pm 0.1 \), known as the ``3D Dainotti" Relation \citep{Dainotti2016}. A similar fundamental plane was found by \cite{Xu2012}, which extends the Dainotti relation with the inclusion of \(E_{\mathrm{iso}}\) rather than $L_{\mathrm{p}, z}$, taking the form: 

\begin{equation}
\log L_X = C_0 + \alpha \log E_{\text{iso}} + \beta \log T^{*}_a
\label{EqLTE}
\end{equation}

\noindent where \( C_0 = 6.0 \pm 2.1 \) (re-parametrized), \( \alpha = 0.86 \pm 0.04 \), and \( \beta = -0.99 \pm 0.04 \), known as the ``$L$–$T$–$E$" correlation  \citep{deng2023pseudo}. 

There have been various debates as to the physical origin of these relations, and whether or not they arise from selection effects. \cite{Band:2005ia} have suggested that the Amati and Yonetoku relation are artifacts of selection biases. \textcolor{black}{More recently, \cite{Heussaff:2013sva} revisited the Amati relation using a homogeneous sample of 43 Fermi/GBM GRBs with known redshift and 243 without, demonstrating that the observed correlation is partly intrinsic—marked by a true paucity of energetic bursts with low $E_{\rm p}$—and partly the result of selection biases that under-detect low-fluence, high-$E_{\rm p}$ outliers and preferentially secure redshift measurements for bursts lying near the lower-right boundary of the $E_{\rm p}$–$E_{\rm iso}$ plane. Further, \citet{2000ApJ...534..227L} applied rigorous non-parametric bias-correction methods to demonstrate that, once selection effects are removed, the intrinsic scatter of the Amati relation increases substantially—though a robust physical correlation remains (See also \citet{dainotti2018pasp130051001} for a detailed review.)
} Analytical models have been able to derive the Amati index as \(E_{\mathrm{p}} \propto E_{\mathrm{iso}}^{0.5}\) under assumptions such as a uniform jet structure with axisymmetric geometry \citep{Eichler2004} or within the cannonball model, which predicts an index of \(1/2 \pm 1/6\) \citep{Dado2012}. For off-axis viewing angles, an index of \(\sim1/3\) has been proposed \citep{Granot2002,ramirez2005}, though recent work suggests that incorporating Lorentz factor effects refines this to a range of \(1/4\) to \(4/13\). Numerical simulations employing both empirical and physically motivated spectra further support these scaling relations, reproducing the observed trends for both on- and off-axis scenarios \citep{Yamazaki2004,Farinelli2021}. While the Yonetoku relation has been primarily interpreted as an on-axis phenomenon linked to the jet's instantaneous emission properties \citep{Zhang2009,Ito2019}, its broader physical interpretation remains an open area of study. 

As for the Dainotti relation, several physical models have been proposed to explain its origin. One leading interpretation involves energy injection from a newly born magnetar, where spin-down due to magnetic dipole radiation sustains the plateau phase and naturally leads to the observed correlation \citep{Rowlinson2014}. Recently, energy extraction from a spinning black hole has also been considered \citep{Lenart2025}. In reality, the situation likely involves a combination of the two.

Regardless of the physical interpretation of these correlations, many have utilized them as distance indicators to derive ``\textcolor{black}{pseudo-redshifts}" of GRBs with no prior redshift measurements \citep{2003A&A...407L...1A, yonetoku2004gamma, 2011ApJDainotti, 2013ApJTan, 2013MNRASTsutsui, 2014ApJ...789...65Y, deng2023pseudo, paul2018modelling}. We find this to be a fundamentally flawed approach. In \cite{Yorgancioglu:2025}, it was shown analytically that the Amati relation yields two redshift solutions, $z_{\mathrm{i}}$, within a reasonable redshift range of $z \in [0.1, 10]$, and hence fails at a fundamental level to serve as a distance indicator. While it was found that the Yonetoku relation can yield a single solution in most cases,  the systematic errors of $z_{\mathrm{i}}$ for individual GRBs are so large that they exceed the characteristic width of the redshift distribution from which they are sampled from. Although it has been acknowledged to some extent that empirical correlations may not be useful for individual GRBs, the assumption remains that they still hold value for population studies. This paper, an extension of \cite{Yorgancioglu:2025}, tests this assumption with the aid of a synthetic catalogue of GRBs. \textcolor{black}{ It should be stressed that the underlying assumption here is that these empirical correlations are intrinsic—i.e. not altered by detector selection effects—to allow direct comparison with prior pseudo-redshift studies. Indeed, the true impact of instrumental biases on these relations remains uncertain; our aim is not to examine those biases but simply to test whether previous pseudo-redshift studies, which necessarily assume no influence from selection effects, provide accurate population inferences.}

In section \ref{methods}, we describe our methodology, in section \ref{results}, we give our results, and in section \ref{Discussion} we give some discussions on future prospects and recommendations for GRB population inferencing.

\section{Methods}
\label{methods}

\noindent The Yonetoku relation is defined in terms of two intrinsic parameters, \(E_{\mathrm{p},z}\) and \(L_{\mathrm{p}, z}\), which are related to their observer frame quantities through 
\begin{equation} 
E_{\text{p},z} = E_{\text{p,o}} \times (1 + z)\,  
\label{eq:intrinsicEp}
\end{equation}
and 
\begin{equation}
L_{\mathrm{p}, z} = 4\pi D_{\rm L}^2 f_{\gamma} k\,
\label{eq:intrinsicEiso}
\end{equation}

\noindent where \(f_{\gamma}\) denotes the observed flux, \(E_{\mathrm{p, o}}\) is the observed peak energy, and \(k\) is the k-correction (which we take as \(k = 1\)), and \(D_{\mathrm{L}}\) is the luminosity distance, i.e.,

\begin{figure}[ht!]
\centering
\includegraphics[width=0.99\linewidth]{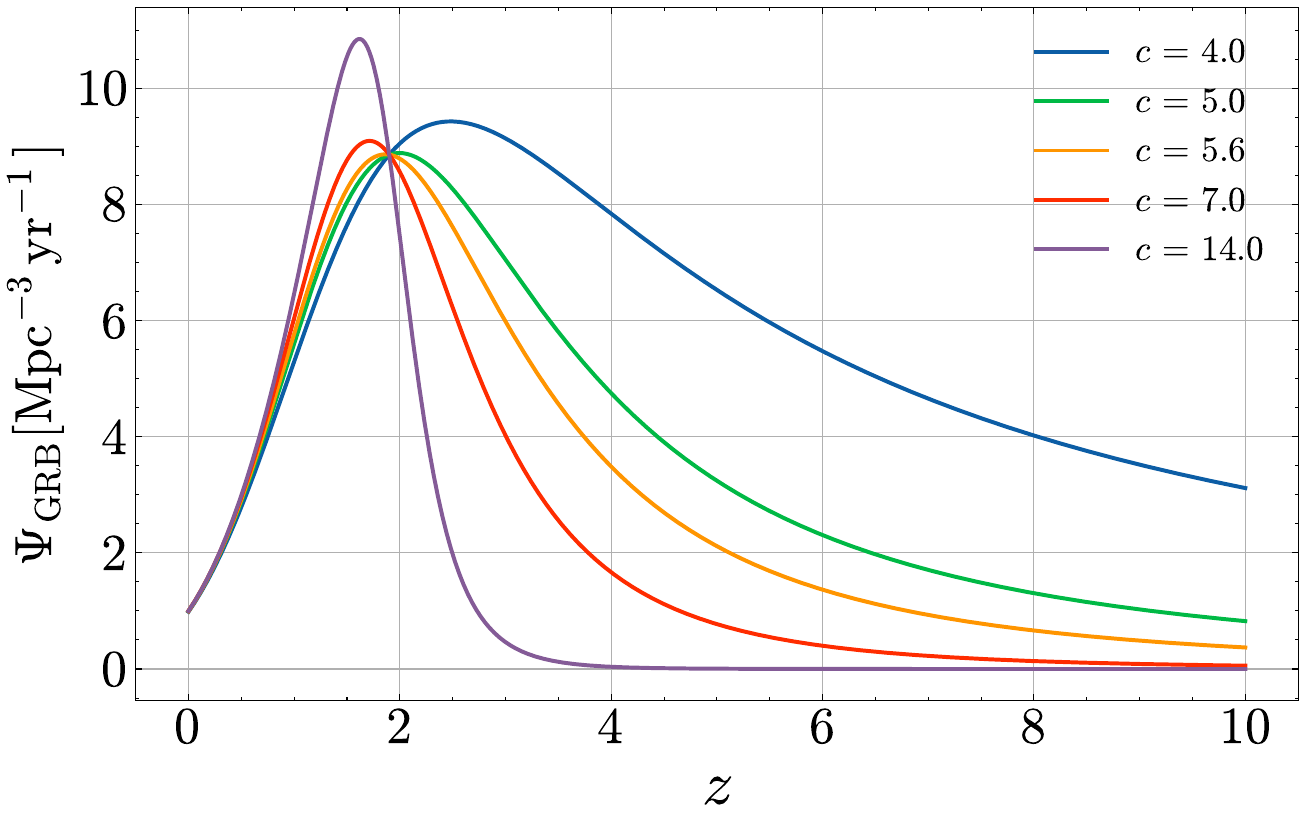} 
\caption{Redshift distribution of the simulated GRB population based on the parametrization of Eq. \ref{eq1}, with varying values of $c$ corresponding to different widths.}
\label{fig:distributions}
\end{figure}

\begin{equation}
D_L(z) = (1 + z) \frac{c}{H_0} \int_0^z \frac{dz'}{\sqrt{\Omega_M (1 + z')^3 + \Omega_\Lambda}}\,.
\end{equation}   

 \noindent We adopt the Planck18 cosmology, assuming a flat $\Lambda$CDM universe with $H_0 = 67.66 \, \mathrm{km} \, \mathrm{s}^{-1} \, \mathrm{Mpc}^{-1}$ and $\Omega_{\mathrm{M}} = 0.3111$ as reported by \cite{aghanim2020planck}. Our methodology follows the prescription of \cite{Yorgancioglu:2025}; for each individual burst with a given \(f_{\gamma}\) and \(E_{\mathrm{p, o}}\), we can trace out a parametric curve defined by equations \ref{eq:intrinsicEp} and \ref{eq:intrinsicEiso}, where each point on this curve corresponds to a redshift between $z \in [0.1, 10]$. We refer to this locus of points as 
 
\begin{equation}
\mathcal{Y}(z; E_{\mathrm{p,o}}, f_{\gamma}) = \left( L_{{\mathrm{p}, z}}(z),\; E_{\mathrm{p},z}(z) \right).
\end{equation}

 \noindent We take the redshift at which there is an intersection between $\mathcal{Y}(z; E_{\mathrm{p,o}}, f_{\gamma})$ and the Yonetoku line as the inferred \textcolor{black}{pseudo}-redshift, $z_{\mathrm{i}}$, where we take the best fit Yonetoku parameters to be $a_{\mathrm{Y}} = 0.625$ and $b_{Y} = -30.22$ \citep{yonetoku2010possible} \footnote{From \cite{yonetoku2010possible}, \[ L_p = 10^{52.43 \pm 0.037} \times \left[ \frac{E_p (1 + z)}{355 \, \text{keV}} \right]^{1.60 \pm 0.082}\] Thus, \(a_{Y}\) = $\frac{1}{1.6}$, and \(b_{Y}\) = $\log(355) - \frac{52.43}{1.6} $ }. To simulate the redshift distribution of GRBs, we assume the following parametrized formulation:

\begin{equation}
    \Psi(z) \propto \frac{(1 + z)^{a}}{1 + \left( \frac{1 + z}{B} \right)^{c}}, 
    \label{eq1}
\end{equation}

where we utilize the best-fit parameters of  \( a = 2.7 \), \( B = 2.9 \), \(c = 5.6\), while also allowing \( c \) to vary to control the characteristic width \citep{madau2014cosmic}. We construct a redshift probability distribution by multiplying Eq \ref{eq1} with the comoving volume element and applying a time-dilation correction:

\begin{equation}
    P(z) = \Psi(z) \frac{dV}{dz} \frac{1}{1 + z}.
    \label{eq2}
\end{equation}

We begin by generating 4000 GRBs \textcolor{black}{(roughly equivalent to the sample size of Fermi-GBM)}, starting with the isotropic peak luminosities, \( L_{\mathrm{p},z} \) by sampling from a lognormal distribution: \( \log_{10} L_{\text{p},z} \sim \mathcal{N}(\mu = 52.5, \sigma^2 = 1) \) \footnote{ \textcolor{black}{All the relevant codes in this paper can be accessed from the Gitlab repository:  \href{https://code.ihep.ac.cn/emre/population-studies}{\texttt{</>}}}}.  Next, we compute the corresponding rest-frame peak energies, \( E_{\mathrm{p},z} \), by assuming that the GRBs follow the best-fit Yonetoku correlation, with an added intrinsic dispersion of \( \sigma_{\log E_{\mathrm{p},z}} = 0.25 \) \citep{ghirlanda2005peak}. Note that the specific form of the luminosity function does not affect the core results, as it only alters the density of sampling within the parameter space; the geometry of the parametric curve \( \mathcal{Y}(z; E_{\mathrm{p,o}}, f_{\gamma}) \), and its intersection with the Yonetoku correlation, remain invariant under this choice.\textcolor{black}{ \ Although our analysis neglects instrumental selection effects due to finite detectability, in Section \ref{Discussion} we demonstrate that imposing a flux-limit is effectively equivalent to adopting a narrower intrinsic population model— which is covered in this work.}

%\textcolor{blue}{We would also like to stress that, although our following analysis is done without applying instrumental selection effects due to finite detectability, as we will show in the discussion section, taking that into consideration by simulating a flux-limited sample is equivalent to using a narrower intrinsic population model, which is covered in our following studies.}

We then generate the redshift values, \( z_{\mathrm{g}} \), by sampling from the normalized redshift probability distribution detailed in Eq. \ref{eq1} and \ref{eq2} , over the interval \( z \in [0.01, 15] \).  Hence, each simulated GRB is characterized by the set of parameters $ \{ L_{\mathrm{p}}, E_{\mathrm{p},z_{\mathrm g}}, f_{\gamma}, E_{\mathrm{p,o}}, z_{\mathrm g} \}$.

We follow the same procedures for the three paramater correlations of 3D Dainotti and $L-T-E$ correlations. In this case, the intrinsic plateau duration $T^{*}_{a}$ is related to its observed counterpart through

\begin{equation}
    T^{*}_{a} = \frac{T^{*}_{a, \mathrm{o}}}{1 + z} 
\end{equation}

Where $T^{*}_{a, \mathrm{o}}$ is the observed plateau duration. In the case of the 3D Dainotti relation, we simulate GRBs following the fundamental plane in Eq \ref{Eq3DD}, with best fit parameters from \cite{Dainotti2016} and dispersion $\sigma_{\rm int} = 0.27$, which is an optimistic lower limit found for a ``gold" sample of 40 GRBs. Each GRB here is defined by the set of parameters $ \{ L_{\mathrm{p}}, T^{*}_{a}, L_{X}, z_{\mathrm g}, f_{\gamma}, T^{*}_{a, \mathrm{o}}, f_{x}\}$, where $f_{x}$ is the obsered plateau flux. The parametric curve takes on the form:

\begin{equation}
 \mathcal{D}(z;  T^{*}_{a, \mathrm{o}},  f_{x},  f_{\gamma}) = \left( L_{{\mathrm{p}, z}}(z),\; T^{*}_{a}(z), \; L_{{X}} (z)\right).
    \label{eq:3ddpc}
\end{equation}

 In the case of the $L$–$T$–$E$ correlation, we simulate GRBs with properties following the fundamental plane as prescribed in Eq \ref{EqLTE}, with dispersion of $\sigma_{\rm int} = 0.36$ \citep{deng2023pseudo}. Each GRB here is then defined by the set of parameters $ \{ E_{\mathrm{iso}}, T^{*}_{a}, L_{X}, z_{\mathrm g}, S_{\gamma}, T^{*}_{a, \mathrm{o}}, f_{x}\}$, where $S_{\gamma}$ is the observed fluence: 

\begin{equation}
E_{\mathrm{iso}} = 4 \pi D_L^2(z) k \cdot \frac{S_{\gamma}}{1 + z}
    \label{eiso}
\end{equation}
The parametric curve then takes on the form: 
\begin{equation}
 \mathcal{L}(z;   S_{\gamma},  T^{*}_{a, \mathrm{o}},  f_{x}) = \left( E_{\mathrm{iso}}(z),\; T^{*}_{a}(z), \; L_{{X}} (z)\right).
    \label{eq:3ddpc}
\end{equation}

\textcolor{black}{Since these relations are recovered from the Swift satellite, which has a narrow energy bandpass, a k-correction approximation of $k=1$ is no longer valid. Assuming a simple power law $N(E) = N_{0}E^{-\alpha}$, we have \citep{2001AJBloom}: }

\textcolor{black}{\begin{equation}
    k = \frac{\int_{E_1/(1+z)}^{E_2/(1+z)} EN(E) \, dE}{\int_{E_1}^{E_2} EN(E) \, dE} = (1+z)^{\alpha -2}
\end{equation}}

\textcolor{black}{Then, $E_{\rm iso}, L_{\mathrm{p}, z}$, and $L_{X}$ now scale as: }

\textcolor{black}{%
\[
E_{\rm iso} 
= 4\pi\,D_{L}^{2}(z)\,\frac{S_{\gamma}}{(1+z)^{3-\alpha_{\gamma}}}
\]
\[
L_{\mathrm{p},z} 
= 4\pi\,D_{L}^{2}(z)\,\frac{f_{\gamma}}{(1+z)^{2-\alpha_{\gamma}}}
\]
\[
L_{X}
= 4\pi\,D_{L}^{2}(z)\,\frac{f_{x}}{(1+z)^{2-\beta_{X}}}
\]
}

\textcolor{black}{To assign values of $\alpha_{\gamma}$ and $\beta_{X}$, we fit a normal Gaussian distribution to the power law indices obtained from the Swift GRB database\footnote{\url{https://swift.gsfc.nasa.gov/archive/grb_table}}.  Specifically, we find: for Swift/BAT: $\alpha_{\gamma} \sim \mathcal{N}(\mu=1.51,\ \sigma=0.49)$, for Swift/XRT: $\beta_{X} \sim \mathcal{N}(\mu=2.03,\ \sigma=0.45)$. We draw $\alpha_{\gamma}$ values from the Swift/BAT distribution to k-correct the prompt-emission quantities ($E_{\rm iso}$ and $L_{p,z}$), and $\beta_X$ from the Swift/XRT distribution to k-correct the afterglow luminosity $L_X$.}

By analogy with Yonetoku, we take the redshift at which the parametric curves $\mathcal{D}(z; T^{*}_{a, \mathrm{o}},  f_{x}, f_{\gamma})$ and $\mathcal{L}(z; T^{*}_{a, \mathrm{o}},     f_{x}, S_{\gamma})$ intersects the corresponding best fit fundamental plane as $z_{\mathrm{i}}$. For all three relations, we gauge whether the distribution of $z_{\mathrm{i}}$ is statistically different than $z_{\mathrm{g}}$ by performing two-sided Kolmogorov-Smirnov (KS) tests \citep{Hodges1958}. We test this for 5 different widths of Eq \ref{eq1} (See figure \ref{fig:distributions}).

\section{Results}
\label{results}

\subsection{Yonetoku Relation}

The normalized histograms and event rates (taken from Eq \ref{eq2}, where \(P(z)\) denotes the unnormalized histogram) for both \( z_{\mathrm{i}} \) and \( z_{\mathrm{g}} \) are presented in Figures~\ref{fig:grb-distributionyon} and~\ref{fig:grb-inferredyon}, for the case of \(c=5.6\). Across all examined values of \( c \), the distribution of \( z_{\mathrm{i}} \) differs significantly from that of \( z_{\mathrm{g}} \), with a reported \( p \)-value effectively equal to zero (see Table~\ref{tab:tab1}). Notably, no clear correlation is observed between the KS statistic and the parameter \( c \). The inferred rate  \( \Psi_{\mathrm{GRB, i}} \) differs significantly from \(\Psi_{\mathrm{GRB, g}}\);  the ratio between \( \Psi_{\mathrm{GRB, i}} / \Psi_{\mathrm{GRB, g}} \) at $z = 0$ is $\sim 2$

In addition, we compared the intrinsic luminosity distribution \( L_{\mathrm{p}, z_{\mathrm{g}}} \) with the inferred distribution \( L_{\mathrm{p}, z_{\mathrm{i}}} \). While the KS statistic in this case is considerably lower (\( \text{KS} = 0.054 \)), the associated \( p \)-value remains statistically significant (\( p = 2.5 \times 10^{-5} \)), indicating that the two luminosity distributions are still highly unlikely to originate from the same parent population.

\subsection{3D Dainotti Relation}

The normalized histograms and event rates for both \( z_{\mathrm{i}} \) and \( z_{\mathrm{g}} \) are shown in Figures~\ref{fig:grb-distribution} and~\ref{fig:grb-inferred}. The KS statistic in this case is higher than that observed in the Yonetoku relation, with an average value of \textcolor{black}{0.24}. Across all tested values of \( c \), the distributions of \( z_{\mathrm{i}} \) and \( z_{\mathrm{g}} \) differ significantly, with corresponding \( p \)-values effectively equal to zero (see Table~\ref{tab:tab1}). This indicates a strong statistical discrepancy between the two redshift distributions. We find that about \textcolor{black}{22\%} of GRBs do not have a solution for \( z_{\mathrm{i}} \), regardless of which intrinsic distribution of $z_{\mathrm{g}}$ the GRBs are sampled from. Moreover, we find a small fraction \textcolor{black}{($\sim$ 6\%)} of GRBs yield two solutions. Significant differences are observed between the intrinsic and inferred GRB rate densities, \( \Psi_{\mathrm{GRB, i}} \) and \( \Psi_{\mathrm{GRB, g}} \), respectively; for instance, at \( z = 0 \), the ratio \( \Psi_{\mathrm{GRB, i}} / \Psi_{\mathrm{GRB, g}} \sim 10\). \textcolor{black}{The KS statistic increases somewhat as $c$ increases. }

We also compared the intrinsic distribution of \( L_X \) with the inferred distribution \( L_{X, z_{\mathrm{i}}} \). The KS statistic in this case was \textcolor{black}{0.14}, with an associated \( p \)-value effectively equal to zero, further confirming a statistically significant deviation between the two luminosity distributions.

\subsection{L-T-E Relation}

\begin{figure*}[ht]
\centering
\hspace*{-0.05\textwidth}
\begin{subfigure}[t]{0.37\linewidth}
    \centering
    \includegraphics[width=\linewidth]{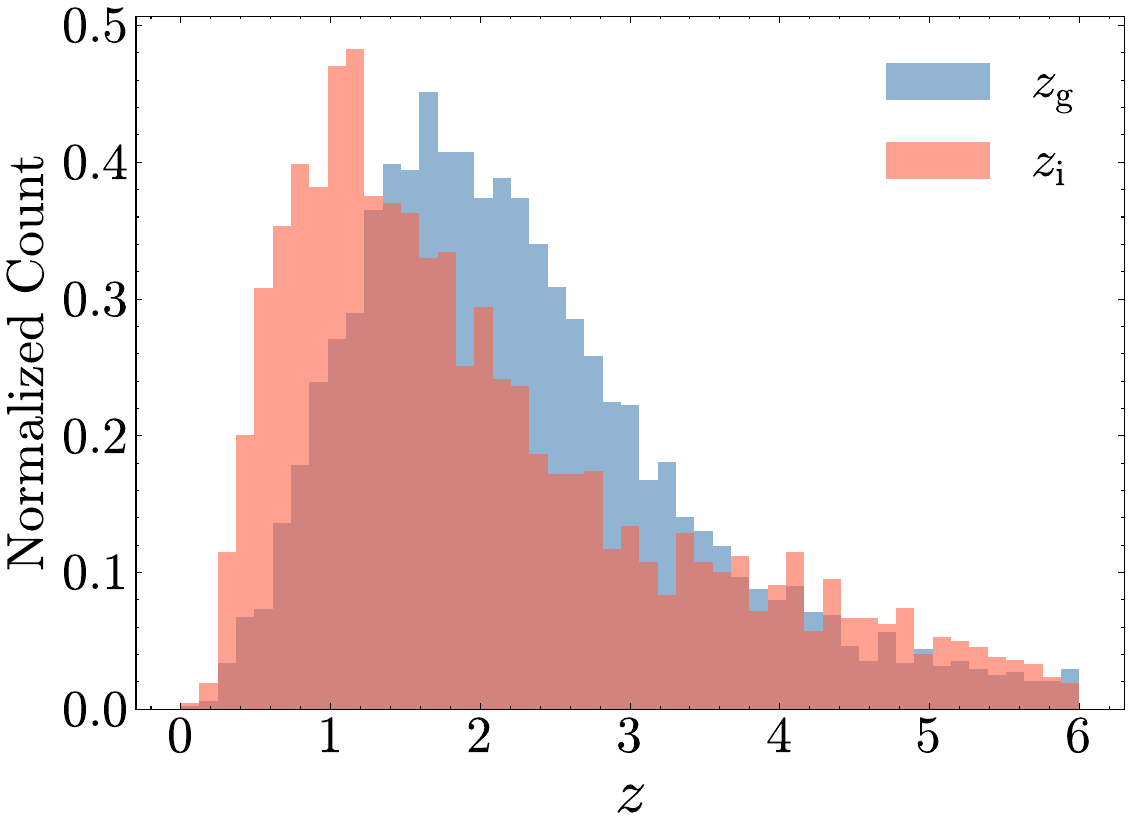}
    \caption{}
    \label{fig:grb-distributionyon}
\end{subfigure}
\hspace{0.00cm}
\begin{subfigure}[t]{0.34\linewidth}
    \centering
    \includegraphics[width=\linewidth]{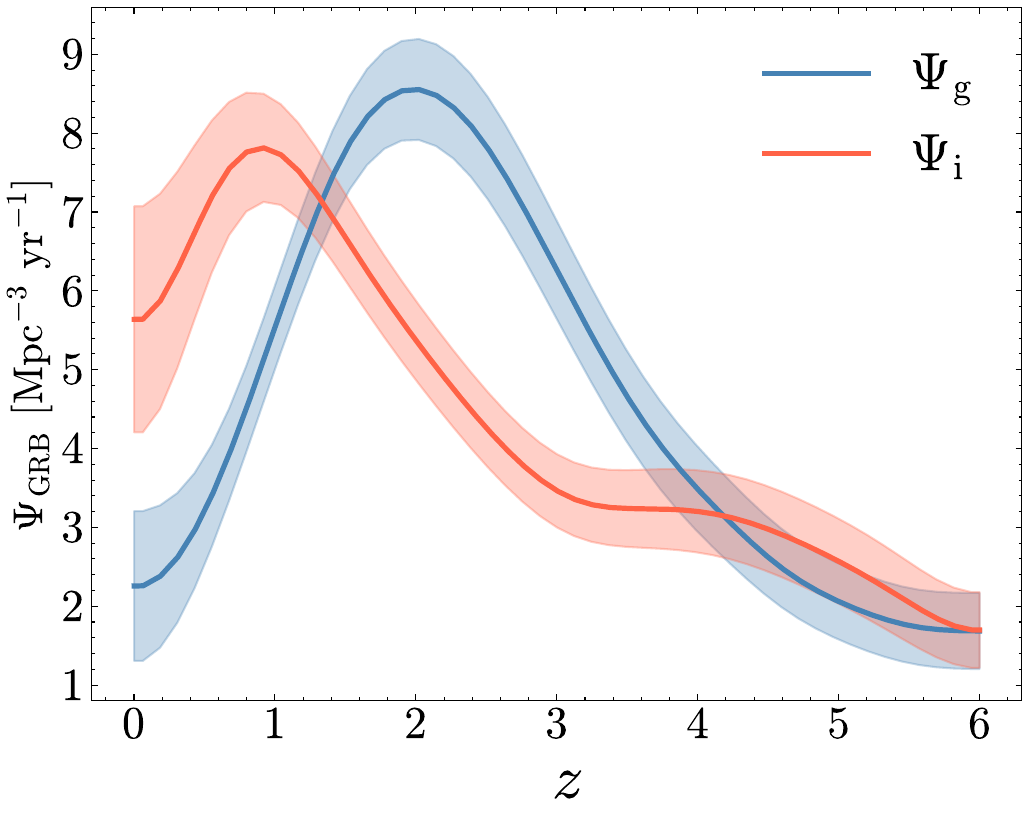}
    \caption{}
    \label{fig:grb-inferredyon}
\end{subfigure}

\label{fig:grb-sidebyside}
\begin{subfigure}[t]{0.37\linewidth}
    \centering
    \includegraphics[width=\linewidth]{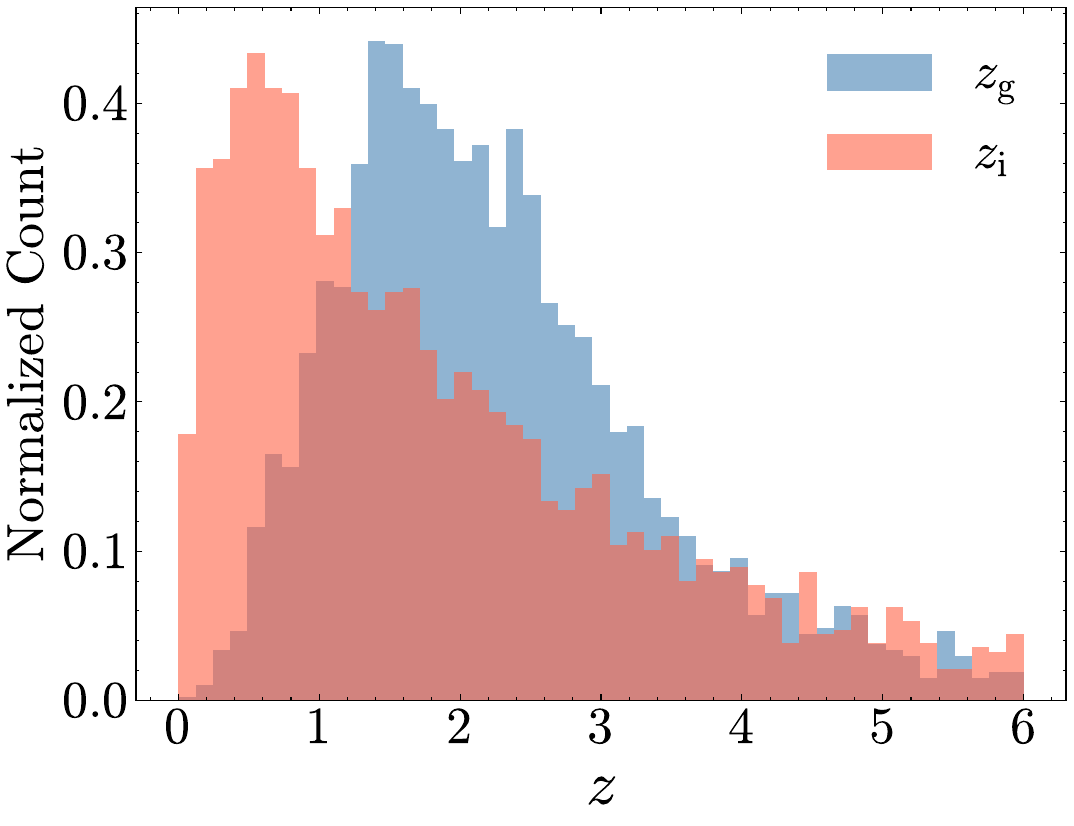}
    \caption{}
    \label{fig:grb-distribution}
\end{subfigure}
%\hfill
\begin{subfigure}[t]{0.39\linewidth}
    \centering
    \includegraphics[width=\linewidth]{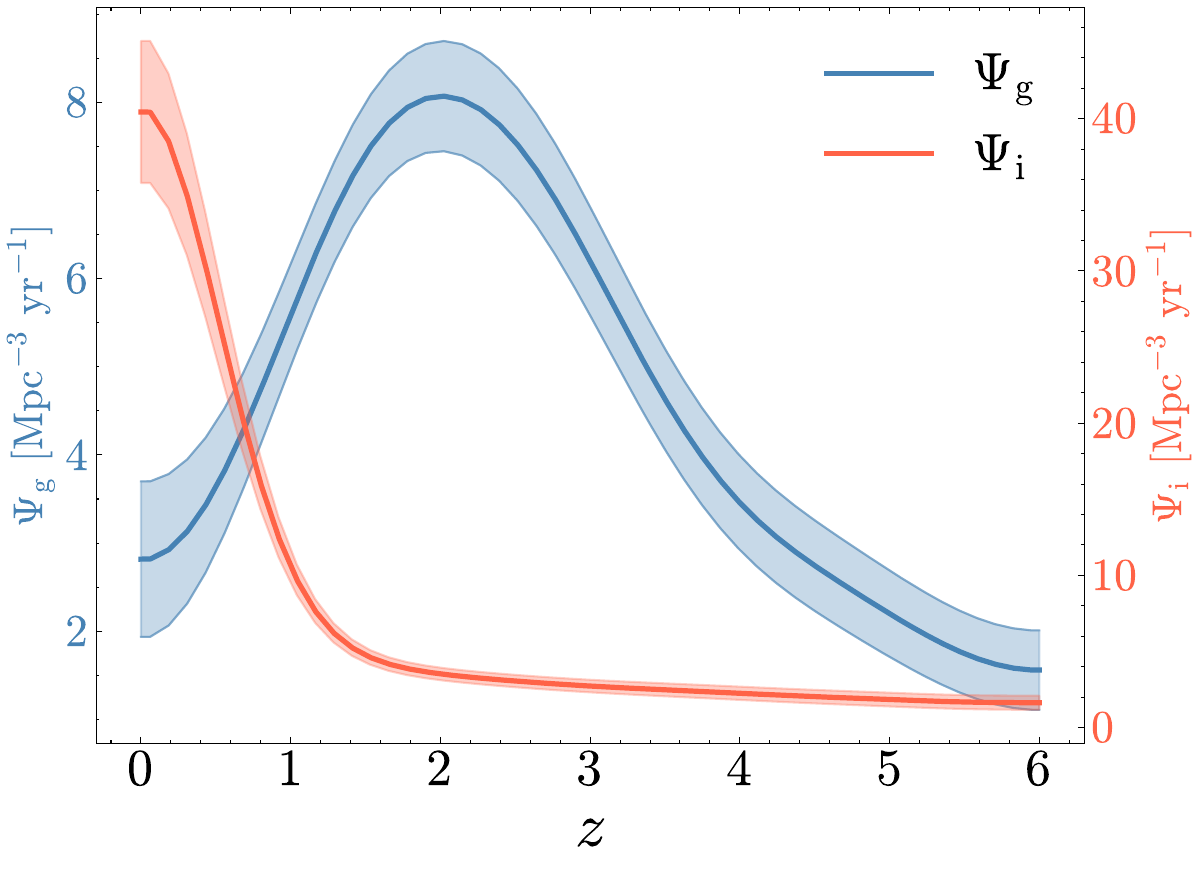}
    \caption{}
    \label{fig:grb-inferred}
\end{subfigure}
%\caption{}
\label{fig:grb1}

\begin{subfigure}[t]{0.37\linewidth}
    \centering
    \includegraphics[width=\linewidth]{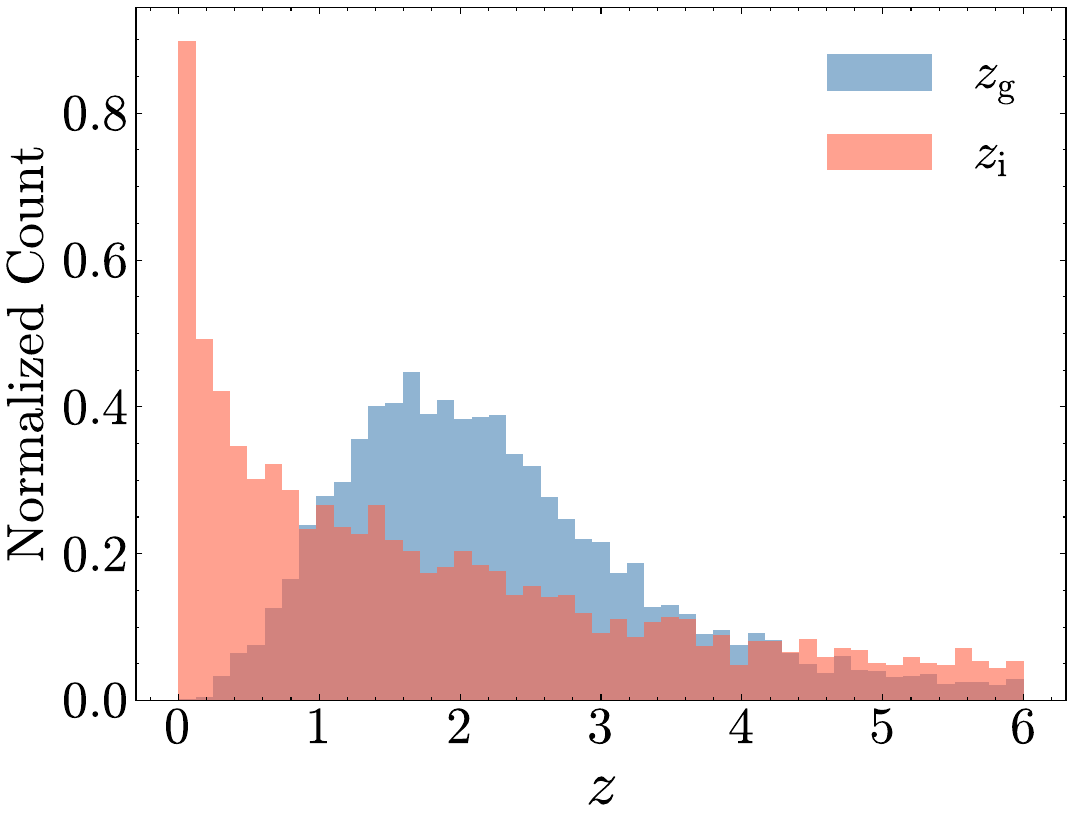}
    \caption{}
    \label{fig:LTE1}
\end{subfigure}
%\hfill
\begin{subfigure}[t]{0.39\linewidth}
    \centering
    \includegraphics[width=\linewidth]{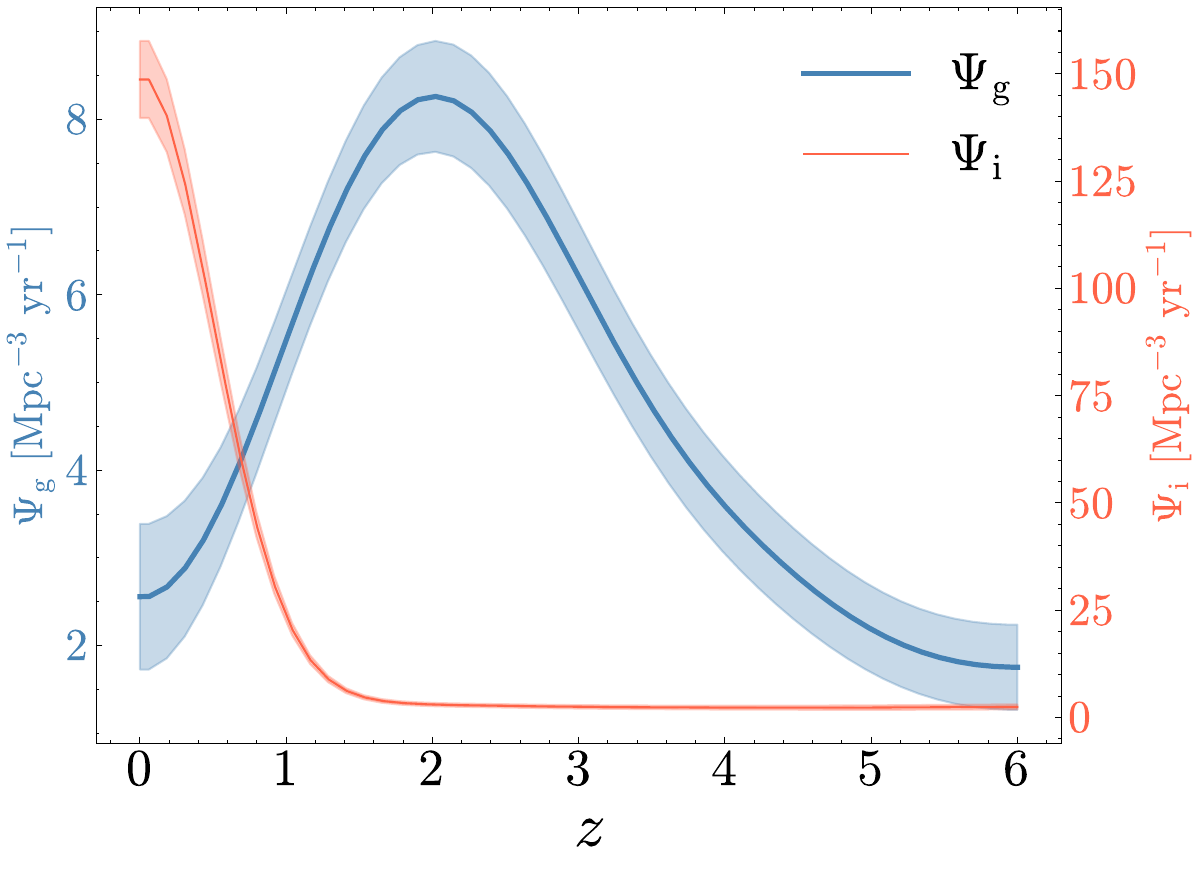}
    \caption{}
    \label{fig:LTE2}
\end{subfigure}
\caption{Panels \textbf{(a)}, \textbf{(c)}, and \textbf{(e)} show the normalized redshift distributions of \( z_{\mathrm{i}} \) (red) and \( z_{\mathrm{g}} \) (cyan) for \( c = 5.6 \), corresponding to the Yonetoku, 3D Dainotti, and $L$–$T$–$E$ relations, respectively. The right-hand panels \textbf{(b)}, \textbf{(d)}, and \textbf{(f)} display the corresponding GRB rate, \( \Psi_{\mathrm{GRB, i}} \)  and \( \Psi_{\mathrm{GRB, g}} \)  (scaled by a factor of $10^{9}$ for visualization) for each case.  The rate densities are smoothed using a Gaussian kernel density estimate. The shaded regions represent the Poisson uncertainty intervals \href{https://code.ihep.ac.cn/emre/population-studies}{\texttt{</>}}. }
\label{fig:grb1}

\end{figure*}

The normalized histograms and event rates for both \( z_{\mathrm{i}} \) and \( z_{\mathrm{g}} \) are shown in Figures~\ref{fig:LTE1} and~\ref{fig:LTE2}. As in the case of the 3D Dainotti relation, the KS statistic in this case is substantially higher than that observed in the Yonetoku relation, with an average value of 0.27. Across all tested values of \( c \), the distributions of \( z_{\mathrm{i}} \) and \( z_{\mathrm{g}} \) differ significantly, with corresponding \( p \)-values effectively equal to zero. \textcolor{black}{We find that 19\% of GRBs yield no solution for $z_{\rm i}$, which is consistent with the findings by \cite{deng2023pseudo}. Moreover, we find that a small fraction ($2.5\%$) yield two solutions. } Significant differences are observed between the intrinsic and inferred GRB rate densities, \( \Psi_{\mathrm{GRB, i}} \) and \( \Psi_{\mathrm{GRB, g}} \), respectively; where at \( z = 0 \), the ratio \( \Psi_{\mathrm{GRB, i}} / \Psi_{\mathrm{GRB, g}} \sim 100 \). \textcolor{black}{The KS statistic increases slightly with increasing \( c \)}.

We also compared the intrinsic distribution of \( L_X \) with the inferred distribution \( L_{X, z_{\mathrm{i}}} \). The KS statistic in this case was \textcolor{black}{0.19}, with an associated \( p \)-value effectively equal to zero.

 % put this in your preamble

\begin{table}[ht]
\centering
\caption{KS statistics and $p$‑values for different values of \(c\) applied to the Yonetoku, 3D Dainotti, and $L$–$T$–$E$ correlations.}
\label{tab:ks-stats}
\begin{tabular}{lccc}
\toprule
\textbf{Relation} & \textbf{$c$} & \textbf{KS Statistic} & \textbf{$p$-value} \\
\midrule
\multicolumn{4}{l}{\textbf{Yonetoku}} \\
& 4.0 & 0.15 & 0.00 \\
& 5.0 & 0.13 & 0.00 \\
& 5.6 & 0.14 & 0.00 \\
& 7 & 0.14 & 0.00 \\
& 14 & 0.20& 0.00 \\
\addlinespace
\multicolumn{4}{l}{\textbf{3D Dainotti}} \\
& 4 & \textcolor{black}{0.23} & 0.00 \\
& 5 & \textcolor{black}{0.23} & 0.00 \\
& 5.6 & \textcolor{black}{0.25} & 0.00 \\
& 7 & \textcolor{black}{0.25} & 0.00\\
& 14 & \textcolor{black}{0.26} & 0.00 \\
\addlinespace
\multicolumn{4}{l}{\textbf{L-T-E}} \\
& 4 & \textcolor{black}{0.26} & 0.00 \\
& 5 & \textcolor{black}{0.26} & 0.00 \\
& 5.6 & \textcolor{black}{0.27}& 0.00 \\
& 7 & \textcolor{black}{0.28} & 0.00\\
& 14 & \textcolor{black}{0.30} & 0.00 \\
\bottomrule
\end{tabular}
\label{tab:tab1}
\end{table}

\section{Discussion \& Conclusion}
\label{Discussion}

Our analysis demonstrates that GRB phenomenological correlations are not only unreliable for estimating redshifts of individual bursts, but also fail to accurately reconstruct intrinsic distributions in population-level studies. Notably, and somewhat counterintuitively, higher-dimensional correlations offer no apparent advantage in analytically deriving \( z_{\mathrm{i}} \). In fact, both the 3D Dainotti and \( L\text{–}T\text{–}E \) relations consistently yield poorer performance than the simpler Yonetoku relation. This is in part due to the high dispersions of the 3D correlations, which in principle is due to several reasons (1) The prompt emission and afterglow originate from distinct physical mechanisms—internal shocks for the prompt phase and external shocks for the afterglow. These phases likely exhibit differing radiative efficiencies, introducing scatter in correlations that combine parameters from both regimes \citep{rees1994unsteady, sari1998spectra, 2000A&A...358.1157D}. The Yonetoku relation relies solely on prompt emission properties, and thus avoids \textcolor{black}{this} cross-phase contamination; (2) Structured jets or variations in jet opening angles between the prompt and afterglow phases could further degrade multi-dimensional correlations. For example, afterglow emission depends on the jet’s lateral expansion and circumburst density, while prompt emission reflects the jet’s initial Lorentz factor and internal dissipation \citep{2023SciA....9I1405O}. If higher-dimensional correlations implicitly assume a uniform jet geometry across phases, systematic errors in inferred parameters (e.g., luminosity) may arise; (3) Afterglow radiation spans radio to X-ray bands, but observational limitations (e.g., Swift-XRT’s 0.3–10 keV bandpass) restrict most studies to partial luminosity estimates. For instance, the absence of robust radio/optical data or high-energy constraints (e.g., NuSTAR observations \(> 79\) keV; \cite{Kouveliotou2013}) makes true bolometric luminosity reconstruction impractical. Correlations relying on X-ray afterglow luminosities are thus prone to underestimating the total energy budget, compounding uncertainties in redshift-dependent parameters.

\begin{figure}[h!]
\centering

\includegraphics[width=0.81\linewidth]{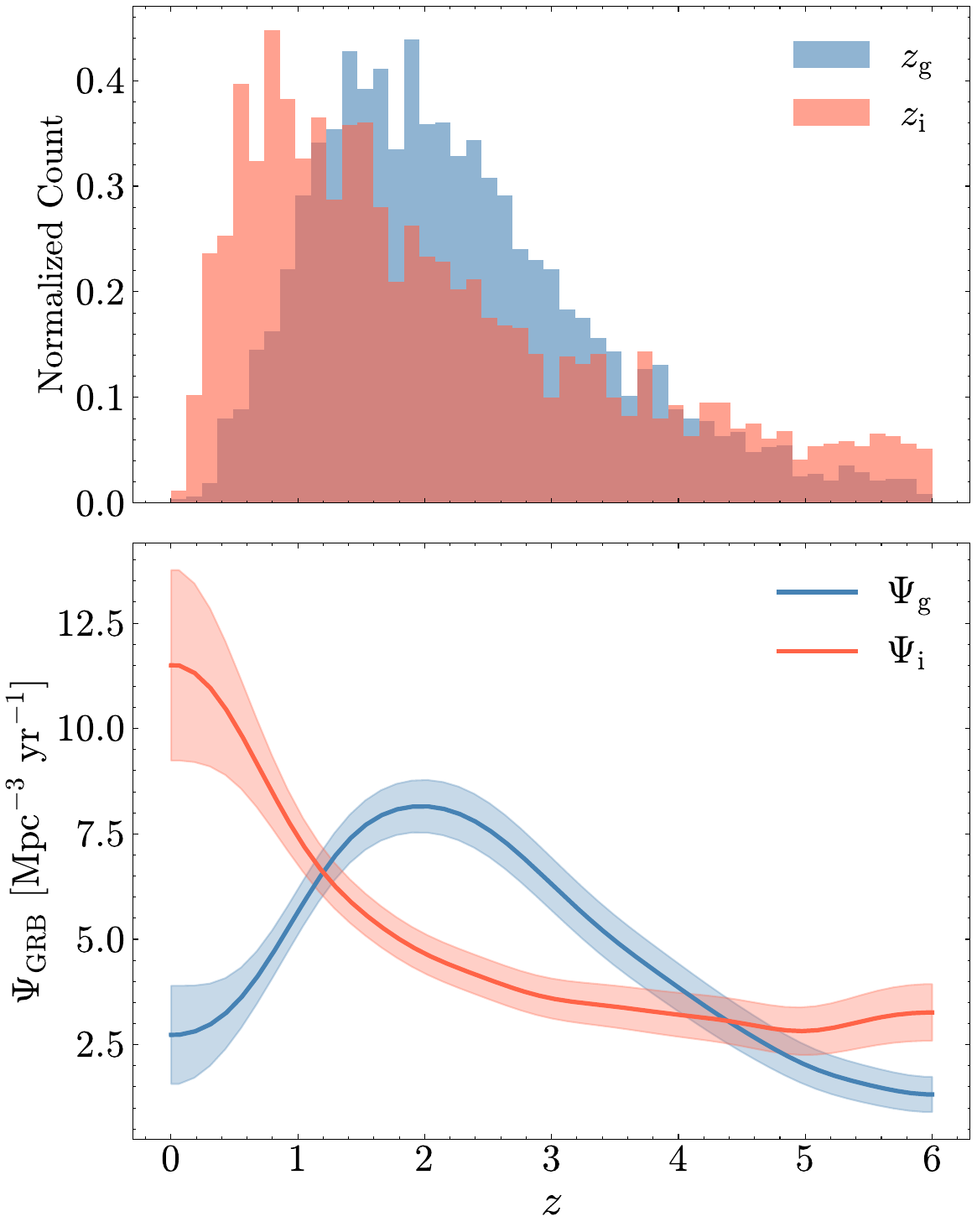} 
\caption{\textbf{Top:} Normalized redshift distributions for the generated redshift \( z_{\mathrm{g}} \) (blue) and the inferred redshift \( z_{\mathrm{i}} \) (red), computed for \( c = 5.6 \). The inferred values are obtained by maximizing the multivariate Gaussian likelihood constructed from the intrinsic 3D Dainotti relation. \textbf{Bottom:} Corresponding GRB rate densities, \( \Psi_{\mathrm{GRB, g}} \) and \( \Psi_{\mathrm{GRB, i}} \), as functions of redshift, scaled by a factor of $10^{9}$ for visualization.}

\label{fig:likelihood}
\end{figure}

\textcolor{black}{In the case of the 3D Dainotti and LTE relation, the inferred redshifts are conspicuously skewed towards low redshifts, $z<1$. } Hence, the inferred redshifts and luminosities are often systematically underestimated, resulting in parametric solutions that intersect the fundamental plane at locations substantially offset from the intrinsic distribution of observed data. \textcolor{black}{We provide a mathematical exposition of this in the appendix.} This discrepancy highlights the inherent limitations of any purely analytical approach to determining \( z_{\mathrm{i}} \), and strongly suggests that redshift inference must instead rely on statistically robust, population-level methodologies. As an example, we compute the mean vector and covariance matrix for the simulated intrinsic distribution of the variables \( L_{\mathrm{p}, z}(z) \), \( T^{*}_{a}(z) \), and \( L_{X} (z) \). These are then used to evaluate the likelihood of each point along the parametric curve derived from the inferred redshift \( z_{\mathrm{i}} \). Specifically, for each redshift-dependent point \( \mathbf{x}(z) = [L_{\mathrm{p}, z}(z),\; T^{*}_{a}(z),\; L_{X}]^{\mathrm{T}} \), we compute the multivariate Gaussian likelihood under the intrinsic distribution:

\begin{align*}
\mathcal{L}(z) &= \frac{1}{(2\pi)^{k/2} |\boldsymbol{\Sigma}|^{1/2}} \\
               &\times \exp\left( -\frac{1}{2} (\mathbf{x}(z) - \boldsymbol{\mu})^{\mathrm{T}} \boldsymbol{\Sigma}^{-1} (\mathbf{x}(z) - \boldsymbol{\mu}) \right)
\end{align*}

where \( \boldsymbol{\mu} \) is the mean vector of the intrinsic distribution, \( \boldsymbol{\Sigma} \) is the corresponding covariance matrix, \( k \) is the dimensionality of the space (in this case, \( k = 3 \)), and \( \mathbf{x}(z) \) denotes the vector of observables as a function of redshift. We take the redshift with the highest likelihood as $z_{\mathrm{i}}$. For the case of \( c = 5.6 \), we obtain a KS statistic of 0.17 with an associated \( p \)-value effectively equal to zero (See Figure \ref{fig:likelihood}). While the result still indicates a statistically significant mismatch and remains unsuitable for reliable inference, it nevertheless marks a notable improvement over the purely analytical method, which gives a KS value of \textcolor{black}{0.25} (see table~\ref{tab:tab1}). We believe that Bayesian hierarchical methods or machine learning, which incorporate detailed spectral shapes, afterglow properties, and prompt light-curve features, could significantly improve pseudo-redshift estimates and population studies. For example, recent works by \cite{aldowma2024deep} and \cite{dainotti2024inferring} have successfully used machine learning to infer GRB redshifts. Their ensemble models leverage a broader set of input variables, such as the low $\alpha$ and high $\beta$ spectral indices of the Band function, allowing them to uncover complex non-linear relationships that traditional regression methods may overlook.  \cite{2020ApJOsborne} adopted a multilevel empirical Bayesian approach to estimate redshifts for 1366 BATSE long-duration gamma-ray bursts (LGRBs). It is no surprise that their redshift estimates differ significantly with the results obtained by \cite{yonetoku2004gamma}, who have utilized the Yonetoku relation to derive \textcolor{black}{pseudo-redshifts}. 

\textcolor{black}{As is evident from Table \ref{tab:tab1}, there is no advantage in using a narrow and ``closer'' injected distribution. Performing the analysis on a flux limited sample of GRBs would thus not yield any benefits. We perform a simple test of this in the case of Yonetoku, assuming a limiting flux of $F_{\rm lim}=2.0\times 10^{-6}$\,erg\,cm $^{-2}$\,s$^{-1}$, which is one order of magnitude \textcolor{black}{higher} than the 64 ms limiting flux of Fermi, $F_{\rm lim}=2.0\times 10^{-7}$\,erg\,cm $^{-2}$\,s$^{-1}$ \citep{2017ApJ...848L..14G}, for the case of $c=5.6$, and obtain a KS-statistic of 0.13, with an effective $p$-value of 0.   }

\begin{acknowledgments}
E.S.Y. acknowledges support from the“Alliance of International
Science Organization (ANSO) Scholarship For Young Talents.” SXY acknowledges support from the Chinese Academy of Sciences (grant Nos. E32983U810 and E25155U110). SNZ acknowledges support from the National Natural Science Foundation of China (Grant No. 12333007 and 12027803). We would like to acknowledge discussions over this topic with Dr. Fiorenzo Stoppa and Rahim Moradi.
\end{acknowledgments}

%\begin{contribution}

%\end{contribution}

%% To help institutions obtain information on the effectiveness of their 
%% telescopes the AAS Journals has created a group of keywords for telescope 
%% facilities.
%
%% Following the acknowledgments section, use the following syntax and the
%% \facility{} or \facilities{} macros to list the keywords of facilities used 
%% in the research for the paper.  Each keyword is check against the master 
%% list during copy editing.  Individual instruments can be provided in 
%% parentheses, after the keyword, but they are not verified.
%\facilities{}

%% Similar to \facility{}, there is the optional \software command to allow 
%% authors a place to specify which programs were used during the creation of 
%% the manuscript. Authors should list each code and include either 

%% citation or url to the code inside ()s when available.
\software{astropy \citep{2013A&A...558A..33A,2018AJ....156..123A,2022ApJ...935..167A}
          }

%% Appendix material should be preceded with a single \appendix command.
%% There should be a \section command for each appendix. Mark appendix
%% subsections with the same markup you use in the main body of the paper.
%%
%% Each Appendix (indicated with \section) will be lettered A, B, C, etc.
%% The equation counter will reset when it encounters the \appendix
%% command and will number appendix equations (A1), (A2), etc. The
%% Figure and Table counter will not reset.

\appendix

\section{Behavior of Solutions}

\noindent \textcolor{black}{We take the solutions to the following plane equations as the inferred redshift, $z_{\rm i}$:}

{\color{black}
\begin{align}
g(z)
&\;=\;-\,\log L_X \;+\; C_0 \;+\; \alpha\,\log E_{\mathrm{iso}}
   \;+\; \beta\,\log T^*_a
\notag\\
&\;=\;-\,\log\!\Bigl(\frac{f_x\,4\pi D_L^2}{(1+z)^{2-\alpha_X}}\Bigr)
   + C_0
   + \alpha\,\log\!\Bigl(\frac{S_\gamma\,4\pi D_L^2}{(1+z)^{3-\beta_\gamma}}\Bigr)
   + \beta\,\log\!\Bigl(\frac{T^*_{a,o}}{(1+z)}\Bigr)
   = 0,
   \label{eq:plane-g}
\\[1ex]
f(z)
&\;=\;-\,\log L_X \;+\; C_0 \;+\; \alpha\,\log L_{\mathrm{p},z}
   \;+\; \beta\,\log T^*_a
\notag\\
&\;=\;-\,\log\!\Bigl(\frac{f_x\,4\pi D_L^2}{(1+z)^{2-\alpha_X}}\Bigr)
   + C_0
   + \alpha\,\log\!\Bigl(\frac{f_\gamma\,4\pi D_L^2}{(1+z)^{2-\beta_\gamma}}\Bigr)
   + \beta\,\log\!\Bigl(\frac{T^*_{a,o}}{(1+z)}\Bigr)
   = 0.
   \label{eq:plane-f}
\end{align}
}

\begin{figure}[H]
  \centering
  % first subfigure: 45% of text width
  \begin{subfigure}[t]{0.45\textwidth}
    \includegraphics[width=\linewidth]{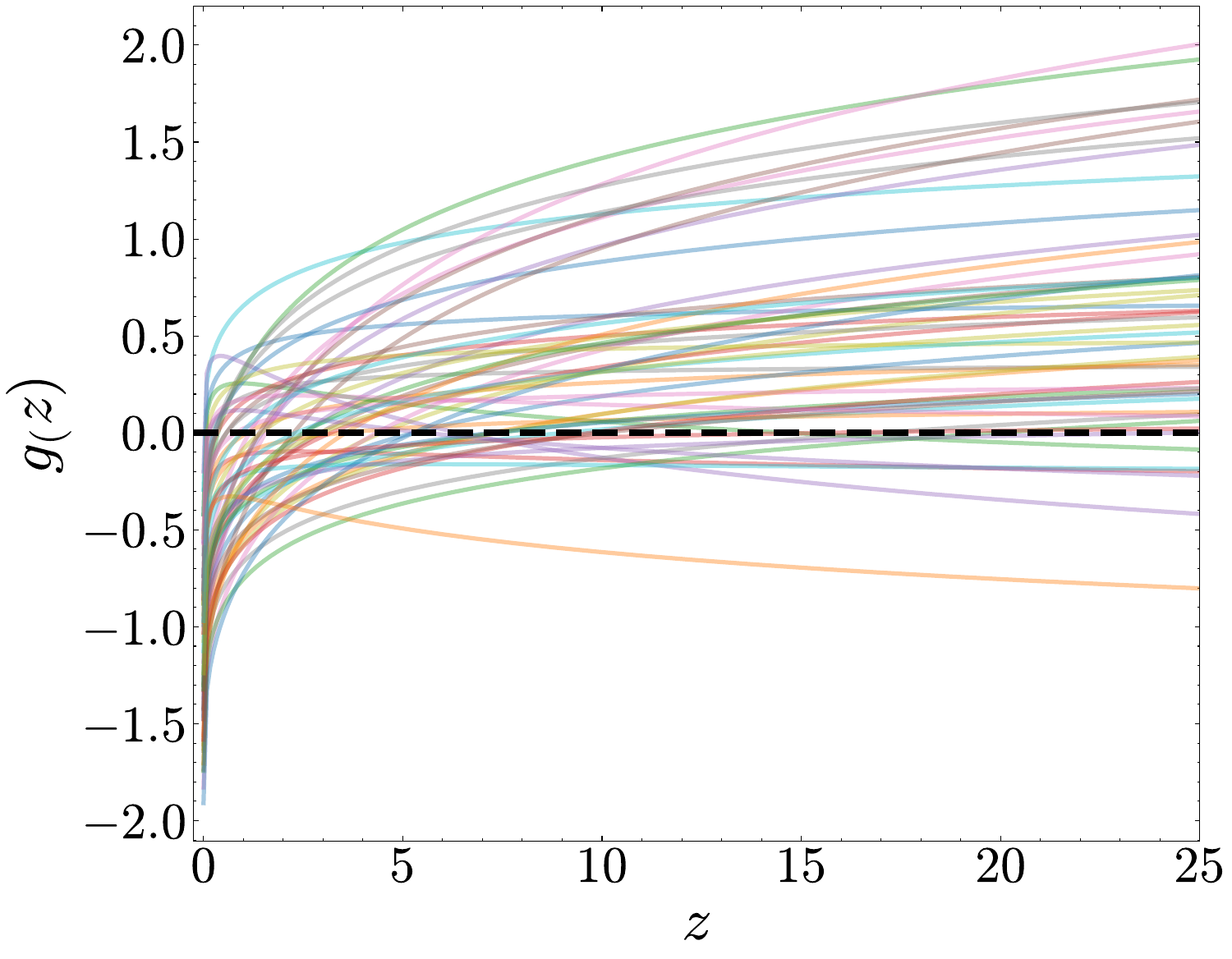}
    \caption{}
    \label{fig:lte1}
  \end{subfigure}%
  \hfill
  % second subfigure: also 45%
  \begin{subfigure}[t]{0.45\textwidth}
    \includegraphics[width=\linewidth]{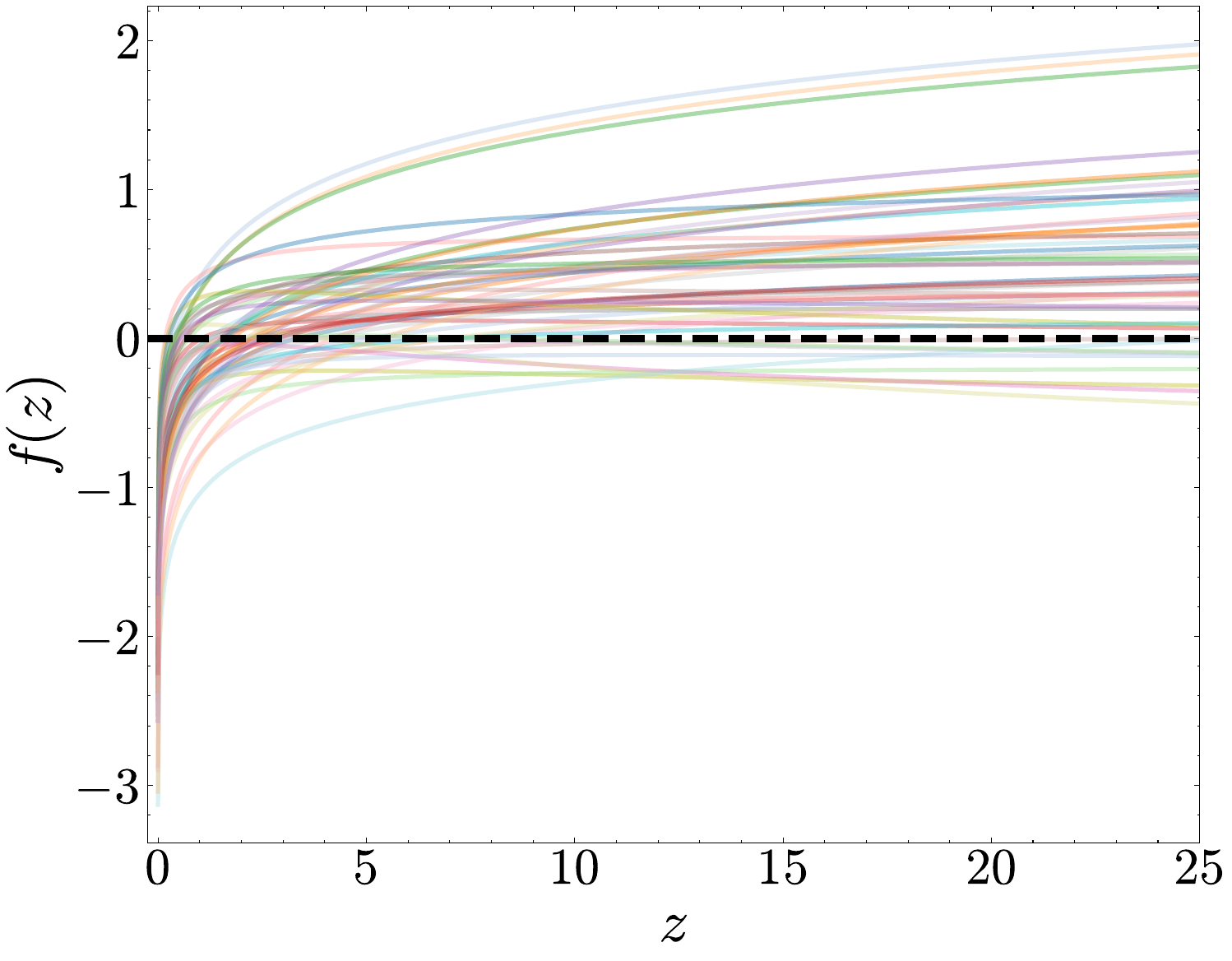}
    \caption{}
    \label{fig:3dd}
  \end{subfigure}

  \begin{subfigure}[t]{0.46\textwidth}
    \includegraphics[width=\linewidth]{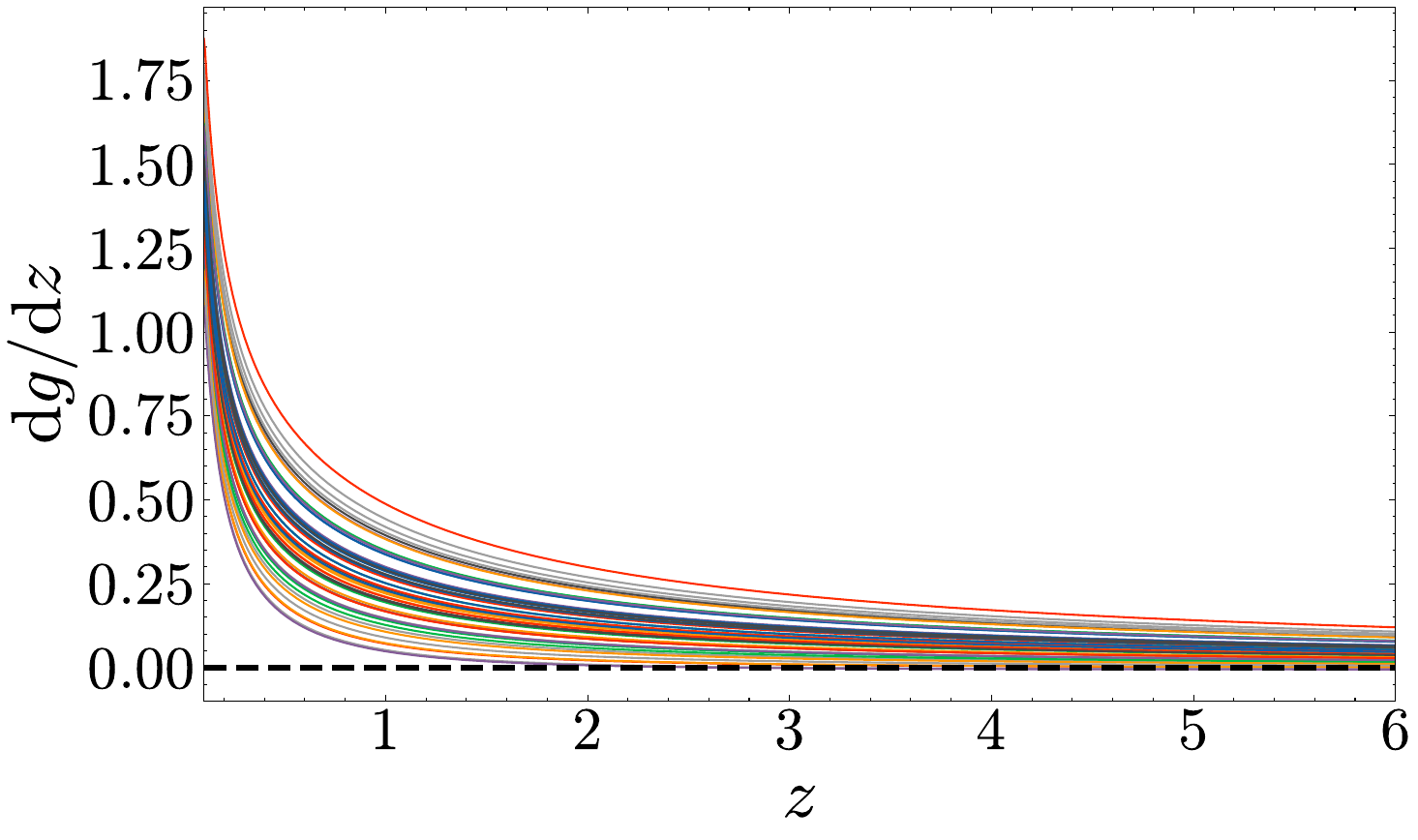}
    \caption{}
    \label{fig:Dlte1}
  \end{subfigure}%
  \hfill
  % second subfigure: also 45%
  \begin{subfigure}[t]{0.46\textwidth}
    \includegraphics[width=\linewidth]{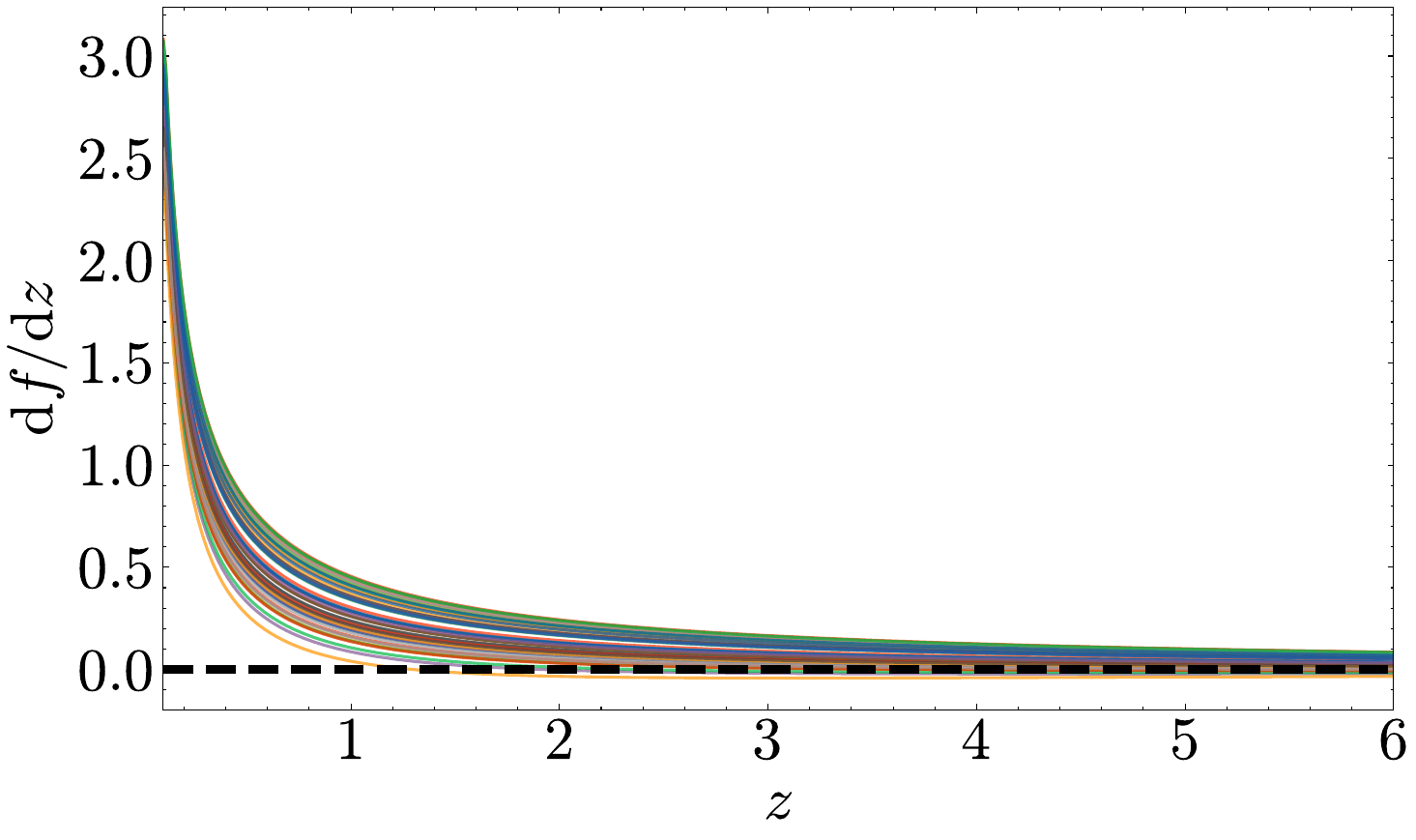}
    \caption{}
    \label{fig:D3dd}
  \end{subfigure}

  \caption{Panels (a) and (b) plot $g(z)$ and $f(z)$ of 50 GRBs, respectively. Their derivatives, $dg/dz$ and $df/dz$ are plotted in panels (c) and (d), respectively \href{https://code.ihep.ac.cn/emre/population-studies}{\texttt{</>}}. }
  \label{fig:comparison1}
\end{figure}

\textcolor{black}{Where $g(z)$ in Eq. \ref{eq:plane-g} corresponds to the plane equation in $L$–$T$–$E$ parameter space, and $f(z)$ in Eq. \ref{eq:plane-f} corresponds to the plane equation in 3D Dainotti space. We plot the curves of $g(z)$ and $f(z)$ for 50 GRBs in figures \ref{fig:lte1} and \ref{fig:3dd}, respectively. In both relations, it is apparent that the curves of $g(z)$ and $f(z)$ have a steep initial rises at low redshifts, which would explain why many inferred solutions are skewed to lower redshifts. To further illustrate this, we plot $dg/dz$ and $df/dz$ in figures \ref{fig:Dlte1} and \ref{fig:D3dd} , respectively. If a GRB does admit a solution ($\exists\,z\colon g(z)=0$), that root always occurs at a redshift smaller than the point where $dg/dz$ first reaches 0; \textcolor{black}{moreover, it is more likely that a solution would occur when $dg/dz$ is large }.  We find that, in the case of both \textcolor{black}{$g(z)$ and $f(z)$}, $dg/dz$ and $df/dz$ $\lesssim 0.5$ \textcolor{black}{when} $z \sim 1$, so many of the roots are likewise confined to this low–$z$ regime.    }

%\subsubsection{Location‐dependence of inferred roots}

\begin{figure}[!hb]
  \centering
  \begin{subfigure}[t]{0.47\textwidth}
    \includegraphics[width=\linewidth]{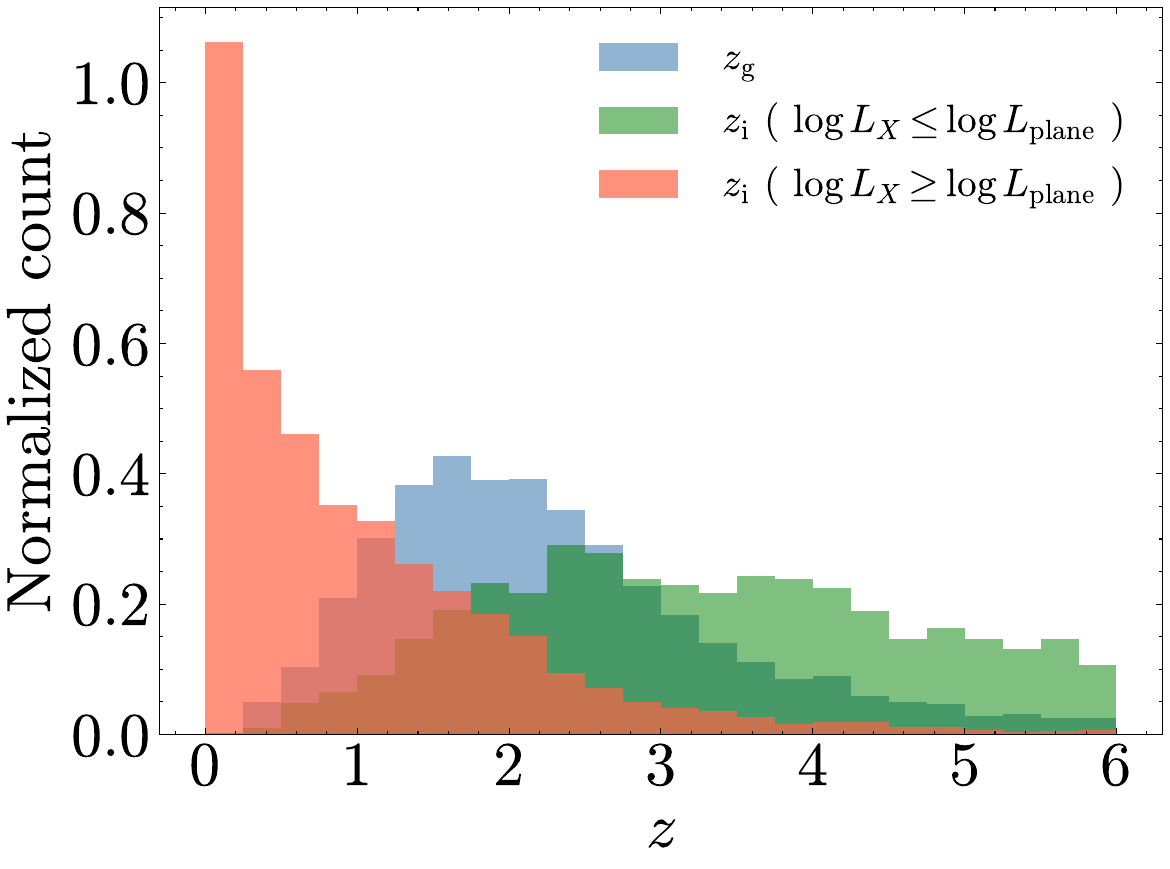}
    %\caption{}
    %\label{fig:abovebelow}
  \end{subfigure}%
  \hfill
  \begin{subfigure}[t]{0.47\textwidth}
    \includegraphics[width=\linewidth]{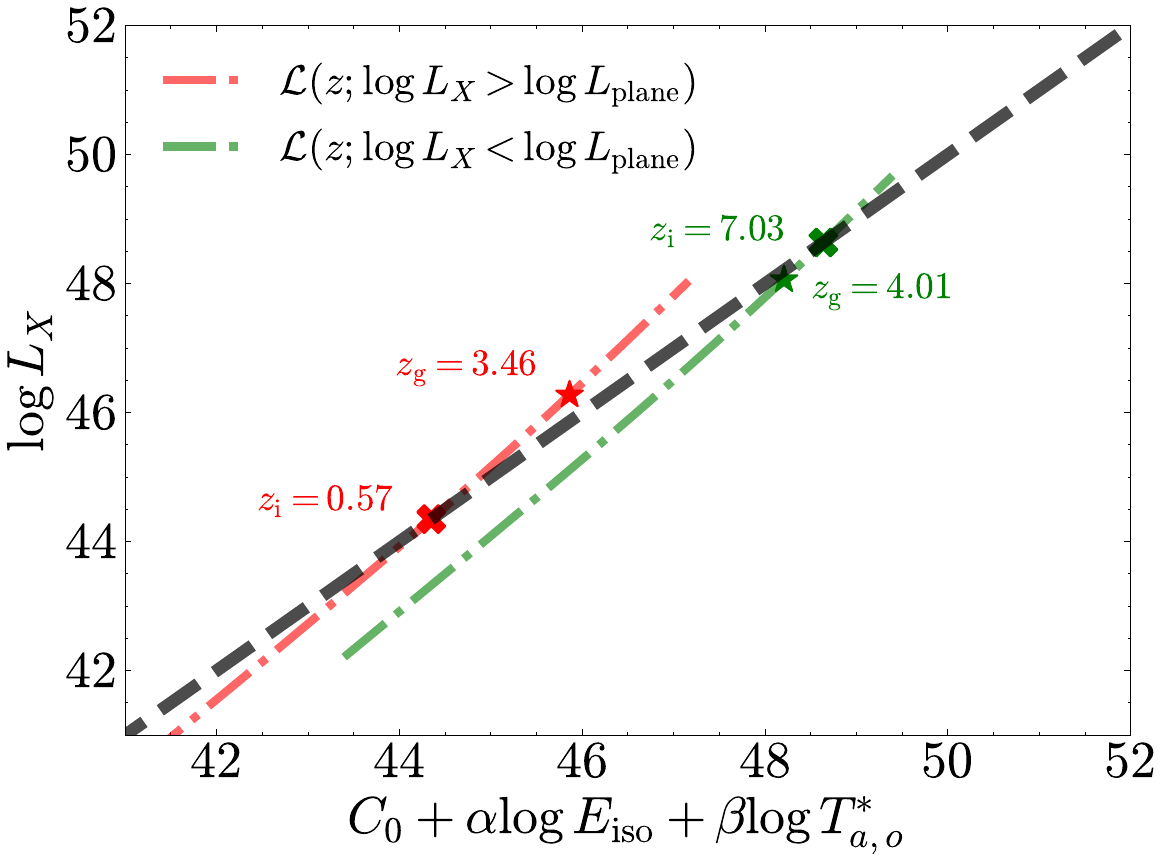}
    %\caption{}
    %\label{fig:ltecurve}
  \end{subfigure}
  \caption{\textbf{Left:} Contributions to $z_{\rm i}$ by GRBs satisfying \(L_X > L_{\rm plane}\) (Red) and \(L_X < L_{\rm plane}\) (Green). \textbf{Right:} Two illustrative curves  $\mathcal{L}(z;   S_{\gamma},  T^{*}_{a, \mathrm{o}},  f_{x})$  satisfying \(L_X > L_{\rm plane}\) (Red) and \(L_X < L_{\rm plane}\) (Green), with $\alpha_{\gamma} = 1.51$ and $\beta_{\gamma}$ = 2.03. The stars denote the true position of the GRB at $z_{\rm g}$, the dashed black line denotes $L_{\rm plane}$, and the $X$ denotes the intersection of $\mathcal{L}(z;   S_{\gamma},  T^{*}_{a, \mathrm{o}},  f_{x})$ with the plane at $z_{\rm i}$ \href{https://code.ihep.ac.cn/emre/population-studies}{\texttt{</>}}.    }
  \label{fig:comparison2}
\end{figure}

\textcolor{black}{\noindent It’s worth noting that the pile‐up at low redshifts arises primarily from those GRBs whose intrinsic positions lie above the best fit plane (i.e. \(\log L_X > \log L_{\rm plane}\), where $\log L_X$ is computed from the true redshift of the GRB, and \( \log L_{\rm plane}\) is computed from the plane defined in Eq.~\ref{Eq3DD} (3D Dainotti space) or Eq.~\ref{EqLTE} ($L$–$T$–$E$ space) evaluated at the same plateau time \(T_a^*\) and isotropic energy \(E_{\rm iso}\) or luminosity $\log L_{\mathrm{p},z}$ of that GRB). We have (in $L$–$T$–$E$ parameter space): 
\begin{equation}
    g(z_{\rm true}) \; \equiv \; \log L_X - \log L_{\rm plane}
    = 
    \begin{cases}
      <0, & L_X < L_{\rm plane},\\
      >0, & L_X > L_{\rm plane}.
    \end{cases}
\end{equation}
GRBs with \(L_X < L_{\rm plane}\) almost always satisfy \(z_{\rm i} > z_{\rm g}\), and vice versa.  In the left panel of Figure~\ref{fig:comparison2}, we distinguish between the contributions to \(z_{\rm i}\) from GRBs above and below the plane in $L$–$T$–$E$ parameter space. Its evident that the overwhelming majority of redshifts for which $z_{\rm i} < 1$ originates from GRBs which satisfy \(L_X > L_{\rm plane}\), as expected. In the right panel of Figure ~\ref{fig:comparison2}, we plot two illustrative parametric curves of $\mathcal{L}(z;   S_{\gamma},  T^{*}_{a, \mathrm{o}},  f_{x})$ in the case of \(L_X > L_{\rm plane}\) and \(L_X < L_{\rm plane}\). 
}

%\noindent \textcolor{blue}{Another way of understanding this, is}

\newpage

%\section{}

%\section{Author publication charges} \label{sec:pubcharge}

%% For this sample we use BibTeX plus aasjournals.bst to generate the
%% the bibliography. The sample7.bib file was populated from ADS. To
%% get the citations to show in the compiled file do the following:
%%
%% pdflatex sample7.tex
%% bibtext sample7
%% pdflatex sample7.tex
%% pdflatex sample7.tex

\bibliography{sample7}{}
\bibliographystyle{aasjournal}

%% This command is needed to show the entire author+affiliation list when
%% the collaboration and author truncation commands are used.  It has to
%% go at the end of the manuscript.
%\allauthors

%% Include this line if you are using the \added, \replaced, \deleted
%% commands to see a summary list of all changes at the end of the article.
%\listofchanges

\end{document}